\documentclass[aps,twocolumn,pra,preprintnumbers,amsmath,amssymb,superscriptaddress]{revtex4}
\usepackage{graphicx}
\usepackage[dvips]{color}
\newcommand{\bea}{\begin{eqnarray}}
\newcommand{\eea}{\end{eqnarray}}
\newcommand{\p}{\partial}

\begin{document}

\title{A rotating hairy AdS$_3$ black hole \\
with the metric having only one Killing vector field}

\preprint{
{\normalsize OU-HET-858}
}


\author{Norihiro {\sc Iizuka}}\email[]{iizuka@phys.sci.osaka-u.ac.jp}
\affiliation{%
{\it Department of Physics, Osaka University, Toyonaka, Osaka 560-0043, JAPAN 
}}

\author{Akihiro {\sc Ishibashi}}\email[]{akihiro@phys.kindai.ac.jp}
\affiliation{%
{\it Department of Physics, Kinki University, Higashi-Osaka 577-8502, JAPAN
}}

\author{Kengo {\sc Maeda}}\email[]{maeda302@sic.shibaura-it.ac.jp}
\affiliation{%
{\it Faculty of Engineering,
Shibaura Institute of Technology, Saitama 330-8570, JAPAN}}

\begin{abstract}
We perturbatively construct a three-dimensional rotating AdS black hole with a real scalar hair. 
We choose the mass of a scalar field slightly above the Breitenlohner-Freedman bound and impose 
a more general boundary condition for the bulk scalar field  at AdS infinity. 
We first show that rotating BTZ black holes are unstable against superradiant modes under our more general boundary condition. Next we construct a rotating hairy black hole perturbatively with respect to a small amplitude $\epsilon$ of the scalar field, up to $O(\epsilon^4)$.  
The lumps of non-linearly perturbed geometry admit only one Killing vector field and co-rotate with the black hole, 
and it shows no dissipation.  
We numerically show that the entropy of our hairy black hole is larger than that of the BTZ black hole with the same energy and the angular momentum. This indicates, at least in the perturbative level, that our rotating hairy black hole in lumpy geometry can be the endpoint of the superradiant instability.  
%
%
\end{abstract}

\maketitle



\noindent


\section{Introduction} 
After the discovery of the gauge-gravity duality~\cite{Maldacena:1997re}, 
Anti-de Sitter (AdS) spacetime has attracted significant interest in both the gravity and high energy physics community. 
Asymptotically AdS space-times exhibit a number of characteristic 
features which are absent in asymptotically flat or de Sitter spacetimes. 
One of such key features is the occurrence of particular types of instability, 
such as weakly turbulent instability~\cite{BizonRost2011} 
and superradiant instability~\cite{CDLY2004}~(see also Ref.~\cite{BritoCardosoPani2015} and references therein), 
in which the AdS boundary plays a crucial role as a confining box. 
It is clearly important to consider consequences -- in particular, 
possible final states -- of such instabilities. 

Consider, for example, superradiant instability of a rotating AdS black hole 
with angular velocity $\Omega$. Suppose a monochromatic wave of 
the form $e^{-i(\omega t-k\varphi)}$ falls into the black hole. 
If $\mbox{Re}[\omega]<k \, \Omega$, then such an incident wave gets amplified 
by the black hole rotation and reflected at AdS infinity towards 
the black hole where it again gets amplified, and the process repeats. 
It was suggested that the end point of this superradiant instability 
may be a rotating black hole which admits only one Killing vector 
field~\cite{KLR2006}. 
In fact, the possibility of such a less symmetric black hole was first 
suggested in \cite{Reall2003} in the context of 
asymptotically flat black holes in higher dimensions.  
Motivated by this, rotating black hole solutions with complex scalar hair 
that break axial symmetry were numerically 
constructed~\cite{DiasHorowitzSantos2011, SPMM, HerdeiroRadu2014}.  
The proposed solutions are, however, not completely satisfactory 
in the sense that although the scalar field configuration is indeed 
invariant only under one Killing vector field, 
the metric itself admits more than one Killing vector field.  

In this paper, we address this issue in a 
three-dimensional setup and construct, for the first time, a rotating AdS black hole whose metric possesses only a single Killing vector field. 
For this purpose, we first consider scalar field perturbation 
of a rotating BTZ black hole~\cite{BTZ}, which is the simplest model of 
rotating AdS black holes \footnote{Compare~\cite{Mann1,Mann2} for some attempts in three-dimensions along the same line 
of \cite{DiasHorowitzSantos2011}.}. 
When the mass-squared of the scalar field takes a certain negative value, 
we can impose, rather than the Dirichlet condition, more general boundary conditions at AdS infinity under which the rotating BTZ black hole exhibits 
instabilities. 
Next, by inspecting quasi-normal modes of the scalar field, we identify 
the marginally stable BTZ solution. 
Then, based on that, we perturbatively construct a rotating AdS black hole 
whose metric possesses only one Killing vector field. 
Interestingly, we find that the total energy of our {\it lumpy} black hole is 
lower than that of the BTZ black hole with the same entropy 
and the angular momentum. This in turn implies that 
our hairy black hole is entropically more favorable than the background 
BTZ black hole with the same mass and angular momentum.

\section{``Superradiant'' instability of rotating BTZ black hole}
Consider the three-dimensional model with Lagrangian 
\begin{align}
\label{action_SC}
L = R+\frac{2}{l^2}-2(\nabla_\mu\phi \nabla^\mu\phi + m^2\phi^2+\eta \phi^4),  
\end{align}
where $l$ represents the AdS scale, $R$ the Einstein-Hilbert term, and  
$\phi$ a real scalar field with some constants $m$  and $\eta$. 
In the absence of the real scalar field, our model~(\ref{action_SC}) admits a rotating BTZ black hole~\cite{BTZ} with the metric 
\begin{align}
\label{metric_BTZ}
& ds^2=-\frac{(r_+^2-r_-^2)^2z}{l^2(1-z)(r_+^2-r_-^2z)}dt^2+\frac{l^2}{4z(1-z)^2}dz^2
\nonumber \\
&+\frac{r_+^2-r_-^2z}{1-z}\left(d\varphi-\frac{r_+r_-(1-z)}{l(r_+^2-r_-^2z)}dt\right)^2, 
\end{align} 
where the outer horizon and the infinity ({\it i.e.,} boundary) are located at $z=0$ and $z=1$ 
as in~\cite{IchiSato94}, and $\varphi$ is the angular coordinate 
with period $2\pi$, and $r_+~(r_-)$ denotes the outer~(inner) horizon 
radius.  

Making the ansatz $\phi=\mbox{Re}[\Pi_1(z)]\cos (\omega t-k\varphi)$ and for the meantime setting $\eta=0$, we find the general solution for $\Pi_1(z)$ 
on the background~(\ref{metric_BTZ}) 
expressed in terms of the hypergeometric function~\cite{IchiSato94}, 
\begin{align}
\label{sol:Pi}
& \Pi_1=C_1 z^{\zeta i}(1-z)^{\frac{1-\sigma}{2}}{}_2 F_1(a,b,c;z)
\nonumber \\
&+C_2z^{-\zeta i}(1-z)^{\frac{1-\sigma}{2}}{}_2 F_1(1+a-c,1+b-c,2-c;z), 
\nonumber \\
& a=\frac{il(l\omega-k)}{2(r_+-r_-)}+\frac{1-\sigma}{2}, \,\,\,
  b=\frac{il(l\omega+k)}{2(r_++r_-)}+\frac{1-\sigma}{2}, \nonumber \\
& c=1+2i\zeta, \,\, \sigma=\sqrt{1+m^2l^2}, \,\, 
\zeta=\frac{l(l r_+\omega -kr_-)}{2(r_+^2-r_-^2)}. 
\end{align} 
Imposing an ingoing boundary condition on the horizon, we set $C_1=0$. 
Then, with the help of \cite{Handbook}, 
the asymptotic behavior is given by 
\begin{align}
\label{asymp_Pi1}
& \Pi_1\simeq {\alpha}(1-z)^{\frac{1-\sigma}{2}} 
              +{\beta}(1-z)^{\frac{1+\sigma}{2}} \,, 
\\
& 
\alpha= \frac{\Gamma(2-c)\Gamma(c-a-b)}{\Gamma(1-a)\Gamma(1-b)} \,, 
\,\, 
\beta= \frac{\Gamma(2-c)\Gamma(a+b-c)}
{\Gamma(1+ a-c)\Gamma(1+b-c)} \,, \nonumber 
\end{align} 
where we have set $C_2=1$. 
The Breitenlohner-Freedman (BF) bound~\cite{BF} corresponds to the case 
$m^2l^2 =-1$. In the range $-1<m^2l^2<0$, both the modes are 
normalizable near the infinity and therefore we can impose 
a more general boundary condition at infinity given by 
$\alpha = \kappa^{-1} \beta$ 
 with $\kappa$ being some constant.  
Such a more general boundary condition corresponds to  
adding a double-trace interaction $\sim (1/2  \kappa^2) \int dx^2 {\cal{O}}^2 $ to the dual field theory~\cite{Witten2001}, where ${\cal{O}}$ is an operator 
dual to $\phi$. 
We will see later that the total energy defined in an appropriate manner~\cite{DHIS2013} 
is indeed conserved. Therefore, in the present context, there is nothing wrong 
with imposing such a more general boundary condition.   

Now we are concerned with the behavior of quasi normal modes, which can 
be computed by fixing the geometric parameters, say $r_-$, $k$, $l$, and 
by imposing the ingoing condition at the horizon and our more general boundary condition at infinity. 
For the purpose of numerical computation below, 
let us set $m^2=-8/9l^2$, $l=1=k$, $r_-=3$, and choose $\kappa$ as 
the following particular value:  
\begin{align}
\label{kappa}
\kappa = - 
\frac{4\sqrt{3}\pi^3  \left(\cosh \frac{\pi }{5} +\frac{1}{2} \right)^{-1}}{ \left( \Gamma\left(\frac{1}{3}\right) \right)^2 \left|\Gamma\left(\frac{1}{3}+\frac{i}{10}  \right)  \right|^{4} } = - 0.414 \,.  
\end{align} 
In this setup, we compute quasi normal modes $\omega=\omega_R+i\omega_I $ 
and plot some of our results in Fig.~1. 

\begin{figure}[ht]
\centering
  \includegraphics[width=8.5truecm,clip]{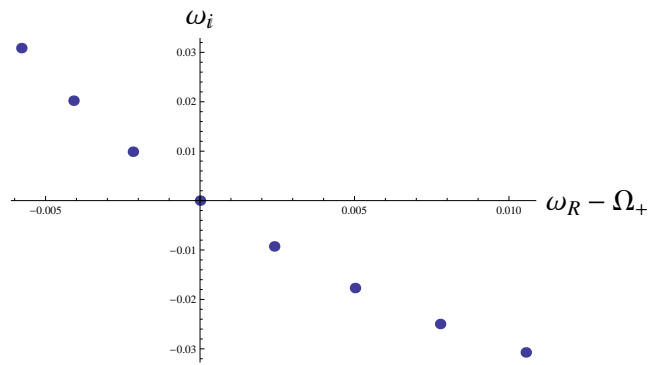} 
\caption{The horizontal line represents $\omega_R-\Omega_+$, while the vertical line represents $\omega_I$. The parameters are chosen as 
$r_-=3$, $k=l=1$, and $\kappa=-0.414$.}
\end{figure}
Note that the outer horizon radius $r_+$ (and thus 
the angular velocity $\Omega_+$) also varies depending on 
the value of $\omega$. In particular, we can find 
that when $r_+=5$, $\omega$ becomes real and equal to $\Omega_+ =3/5$ 
as shown in Fig.~1, and also that there is no flux across the horizon. 
In fact, to achieve this, we have purposely chosen the particular value of 
$\kappa$ in Eq.~(\ref{kappa}). 
Thus the rotating BTZ black holes become unstable under our more general 
boundary condition against modes with $\omega_I>0$ shown in Fig.~1. 
In particular, the solution with $r_+=5$ and $r_-=3$ can be viewed 
as the marginally stable solution. 
This is not the case under the standard Dirichlet boundary 
conditions~\cite{Birmingham}(see also \cite{StotynMann}). 
We also note that the BTZ background admits a Killing vector field which is 
causal everywhere outside the event horizon, and therefore there would not 
be superradiant instability in the standard sense, 
according to the argument of \cite{HawkingReall}. 
However note also that our model violates the dominat energy condition 
and thus the agument of \cite{HawkingReall} does not straightforwardly 
apply to the present case. 
Since our unstable modes $\omega_I>0$ appear only when $\omega_R<\Omega_+$, 
i.e., the standard superradiant condition is apparently satisfied 
as can be seen in Fig.~1, we loosely call our instability ``superradiant instability'' in the rest of the paper. 

\section{Perturbative construction of a hairy black hole in lumpy geometry} 

Having obtained the marginally stable solution, we can expect that 
there should exist a hairy black hole solution dressed 
with a condensed real scalar field. 
We perturbatively construct such a hairy black hole solution by expanding 
the metric functions and scalar field in a series of small 
amplitude $\epsilon$ of the scalar field, up to $O(\epsilon^4)$, 
starting from our marginal solution.

In the probe limit of the scalar field, where the backreaction onto the spacetime is ignored, 
$\phi$ depends on the coordinate $(r_-/r_+ l) t-\varphi$ and $z$ only, we expect 
that the resultant hairy black hole also depends only on 
$y=\omega_* t-\varphi$ and $z$. 
Thus, we make the metric ansatz as 
\begin{align}
\label{metric_z}
& ds^2=-f(y, z)e^{-2\delta(y,z)}dt^2+\frac{g'(z)^2dz^2}{4g(z)f(y, z)} \nonumber \\
&+g(z) (d\varphi-\Omega(y, z)dt)^2 \,, 
\end{align} 
where $f=0$ on the horizon $z=0$. 
The scalar field $\phi$ and the metric functions, collectively denoted 
by $F$, can be expanded as 
\begin{align}
\label{expansion}
& \phi(y,z)=\epsilon\phi_1(y,z)+\epsilon^3\phi_2(y,z)+\cdots, \nonumber \\
& F=F_0(y,z)+\epsilon^2F_1(y,z)+\epsilon^4F_2(y,z)+\cdots, \nonumber \\
& \omega_*=\frac{3}{5l}+\epsilon^2\omega_1+\cdots,  \,\,\, F=f,\,g, \,\Omega,\,\delta,  
\end{align}
where $F_0$ and the leading term of $\omega_*$ respectively 
denote the corresponding values of the marginal BTZ black hole. 
(Remember that we have set $r_+ = 5$, $r_- = 3$, $l=1$.) 
We require the following asymptotic conditions for the metric functions,  
\begin{align}
\label{asymp_bc}
\lim_{z\to 1}\Omega = \lim_{z\to 1}\delta =0 
\end{align}
and the condition~(\ref{kappa}) for $\phi$.  
Near the horizon, the scalar curvature $R$ becomes 
$R\sim (\omega_*-\Omega)^2(\p_y \phi)^2/f$. 
To remove the singularity on the horizon, we require 
\begin{align}
\label{horizon_bc}
\omega_*= \Omega \big|_{z=0}, 
\end{align}
where $\Omega \big|_{z=0}$ is in fact independent of $y$ as well, 
thus a true constant as can be seen in Eq.~(\ref{ansatz_F1}) below. 
The solution for $\phi_1$ is given by $\phi_1=\Pi_1(z)\cos(ky)$ and using (\ref{horizon_bc}), 
the equation of motion for $\Pi_1(z)$ reduces to  
${\cal L}_k\Pi_1=0$ with ${\cal L}_k$ being the $k$-dependent 
differential operator 
\begin{align}
\label{Eq:Pi_1}
{\cal L}_k := z\p_z^2+\p_z-\frac{l^2(m^2r_+^2+k^2(1-z))}{4r_+^2(1-z)^2}.   
\end{align}
Note that the solution to this equation corresponds to Eq.~(\ref{sol:Pi}) with $\eta =0$. 
From the structure of the Einstein equations, $F_1(y,z)$ can be expanded as 
\begin{align}
\label{ansatz_F1}
& f_1(y,z) \, = \,z(P(z)\cos (2ky)+Q(z))+a_1f^{(1)}_{\rm btz}(z), \nonumber \\
& \delta_1(y,z) \,= \,R(z)\cos (2ky)+S(z),  \,\,\, g_1(z)=a_1g^{(1)}_{\rm btz}(z), \nonumber \\
& \Omega_1(y,z)=zT(z)\cos (2ky)+U(z),
\end{align}
where $f^{(1)}_{\rm btz}(z)$ and $g^{(1)}_{\rm btz}(z)$ are trivial perturbed solutions \footnote{Here we mean that $f^{(1)}_{\rm btz}(z)$ and $g^{(1)}_{\rm btz}(z)$ are solutions of source-free differential equations for metric perturbation.} of the BTZ black hole~(\ref{metric_BTZ}) such that 
$r_+\to r_++a_1\epsilon^2+a_2\epsilon^4+\cdots$ and $r_-\to r_-+a_1r_-/r_+\epsilon^2+\cdots$. Note that this deviation 
does not change the velocity of the black hole. Then, the parameter $a_1$ is not 
included in the perturbed functions $(P,Q,R,S,T,U)$.  

The functions $(P,\,R,\,T)$ representing an oscillating mode are decoupled from the zero mode functions $(Q,\,S,\,U)$ at $O(\epsilon^2)$. 
From the $tz$, $\varphi z$, and $\varphi\varphi$ components of the Einstein equations, we obtain 
a master equation for the oscillating mode 
\begin{align}
\label{Eq:Sec_T}
& zT''+\left(4+\frac{2k^2l^2}{r_+^2-r_-^2}-\frac{2r_+^2}{r_+^2-r_-^2z}   \right)T'
\nonumber \\
&-\frac{2r_-^2(r_+^2-r_-^2+k^2l^2)}{(r_+^2-r_-^2)(r_+^2-r_-^2z)}T
=S_T(\Pi_1, \Pi_1').  
\end{align}  
$P$ and $R$ are expressed by $T$, $T'$, $\Pi_1$, and $\Pi_1'$ only [See 
Appendix]. 

Similarly, we obtain the equations of motion for the zero mode and find 
the regular solutions satisfying (\ref{asymp_bc}) as 
\begin{align}
\label{Eq:Q}
& h(z)=-\frac{k^2 r_-^2}{r_+^2}\int^z_0 \frac{\Pi_1^2}{1-z}dz
-\frac{4r_+^2}{l^2}\int^z_0 (1-z)\Pi_1'^2 dz \nonumber \\
&+\frac{2r_+r_-}{l}\left[U(z)-U(0)   \right]
-\frac{2r_+^2(r_+^2-r_-^2)}{l^2}\left[\frac{S(z)}{r_+^2-r_-^2z}-\frac{S(0)}{r_+^2}\right], 
\nonumber \\
&S=\frac{1}{r_+^2-r_-^2}\int^1_z\left[\frac{k^2r_-^2l^2}{2r_+^2}\Pi_1^2+2(1-z)(r_+^2-r_-^2z)\Pi_1'^2   \right]dz, 
\nonumber \\
& U= \frac{r_+^4(z-1)B}{(r_+^2-r_-^2z)(r_+^2-r_-^2)} \nonumber \\ 
& + \frac{z-1}{(r_+^2-r_-^2z)(r_+^2-r_-^2)} 
  \int^z_0(r_+^2-r_-^2z')^2{\cal S}_U(z')dz', 
\nonumber \\
& 
+\frac{1}{r_+^2-r_-^2} \int^1_z (r_+^2-r_-^2z')(z'-1){\cal S}_U(z')dz',  
%
%
 \\
& {\cal S}_U=-\frac{r_-lk^2}{2r_+(1-z)(r_+^2-r_-^2z)}\Pi_1^2-\frac{2r_+r_-(1-z)}{l(r_+^2-r_-^2z)}\Pi_1'^2,  \nonumber
\end{align}
where $h=zQ+\frac{2(r_+^2-r_-^2)z}{l^2}\Pi_1\Pi'_1$ and $B$ is an integration constant. For simplicity, setting $B=0$, we easily find that 
\begin{align}
\label{h_asymp}
h(1)=2\Omega_+(r_+^2-r_-^2)U'(1), \quad  \Omega_+=\frac{r_-}{r_+l}. 
\end{align}
As shown later, this relation agrees with the first law of our 
hairy black hole.

\section{Higher order solutions for $\epsilon$}
The equations of motion for $\phi_2(y,z)$ are given by 
\begin{align}
\label{Eq:sec_phi}
& \phi_2(y,z)=\phi_{21}(z)\cos (ky)+\phi_{23}(z)\cos (3ky), \nonumber \\
& {\cal L}_k\phi_{21}={\cal S}_{\phi 1}, \quad {\cal L}_{3k}\phi_{23}={\cal S}_{\phi 3},
\end{align}
where the source terms ${\cal S}_{\phi 1}$ and ${\cal S}_{\phi 3}$ include the metric functions for $O(\epsilon^2)$, $\Pi_1$, and 
the parameters $a_1$, $B$. Here we simply omit the precise expression of 
those source terms, as their explicit form is rather complicated and 
not essential for the rest of our arguments. 
Here, $\omega_1$ is replaced by $U(0)$ by Eq.~(\ref{horizon_bc}). 
Again just for simplicity, we hereafter set $B=0$. The second equation can be 
formally solved by constructing two independent homogeneous solution $\lambda_i~(i=1,2)$ 
satisfying ${\cal L}_{3k}\lambda_i=0, \,i=1,2$, where $\lambda_1$ and  $\lambda_2$ are solutions 
satisfying our more general boundary condition~(\ref{kappa}) at infinity and 
the regularity at the horizon. 
The explicit form of $\phi_{23}$ is given in Appendix. 
  
The solution for $\phi_{21}$ in (\ref{Eq:sec_phi}) cannot be obtained 
by using a similar manner of the case $\phi_{23}$.   
This is because there is no regular homogeneous solution, independent of $\Pi_1$~\footnote{Note that $\Pi_1$ is the homogeneous 
regular solution satisfying our more general boundary condition~(\ref{kappa})}. 
By shooting the parameter $a_1$, however, 
we can numerically find the regular solution satisfying our more general boundary condition. Fig.~2 shows the numerical data for 
$\phi_{23}$ and $\phi_{21}$ for $r_+=5$, $r_-=3$, $k=1$, $l=1$, and $\eta=1$. $a_1$ is approximately $\simeq 473$. 
The asymptotic functions of $\phi_{21}$ and $\phi_{23}$ are well approximated 
as $\phi_{21}\simeq -4.31(1-z)^{\frac{1}{3}}+1.79 (1-z)^{\frac{2}{3}}$ and 
$\phi_{23}\simeq -12.5(1-z)^{\frac{1}{3}}+5.18(1-z)^{\frac{2}{3}}$ near infinity, agreeing with our more general boundary condition~(\ref{kappa}). 
\begin{figure}[ht]
\centering
  \includegraphics[width=7truecm,clip]{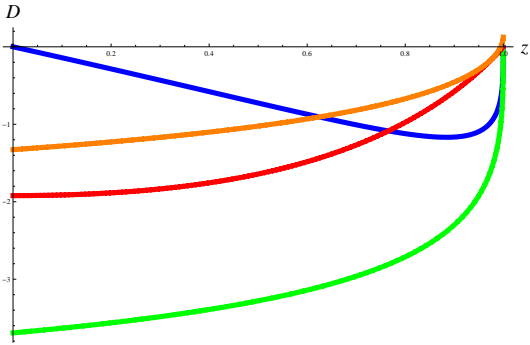} 
\caption{$D=\phi_{21}$ (solid, blue), $D=\phi_{23}/2$ (dashed, green), and $D=\Omega_{20}$ (dotted, red), 
$D=10^{-3} \cdot h_{20}$ (dotdashed, orange) for the parameter choice $r_+ = 5$, $r_-=3$, $k=1$, $l=1$, and $\eta=1$.}
\end{figure}

Similarly, we also numerically construct the metric functions for $O(\epsilon^4)$. 
Since we are interested in the energy difference between the hairy BH and BTZ solution, we only construct the zero 
mode solution~($F_2$ solutions which are independent of $y$) at $O(\epsilon^4)$. Denoting $\Omega_{20}$, $f_{20}$ as 
the zero mode solutions of $\Omega_2$, $f_2$,  
Fig.~1. shows the functions of 
$\Omega_{20}$ and $h_{20}$ for the same parameter choice, where $h_{20}$ is defined by
\begin{align}
\label{h20}
h_{20}(z):=f_{20}+\frac{2(r_+^2-r_-^2)}{l^2}\left(
\Pi_{1}\phi_{21} + \frac{a_1}{r_+} \Pi_1^2 \right)'. 
\end{align}
Note that $h_{20}$ becomes finite at $z\to 1$ since the divergent 
term in $f_{20}$; 
\begin{align}
f_{20}\simeq \frac{4(r_+^2-r_-^2)}{3r_+l^2}{\alpha}(a_1{\alpha}+r_+\alpha_2)(1-z)^{-\frac{1}{3}}, 
\end{align}
cancels with the second divergent term in (\ref{h20}), where $\alpha_2$ is the asymptotic leading 
coefficient of $\phi_{21}\simeq \alpha_2(1-z)^{1/3}$.

\section{Calculation of effective holographic energy}
Following the re-normalization method~\cite{BalaKraus1999} and \cite{DHIS2013}, we calculate the 
effective holographic energy momentum tensor 
\begin{align}
\label{holographic_tensor}
& T_{ij}=K_{ij}-K h_{ij}-\frac{1}{l} h_{ij} \nonumber \\
&\,\,\quad -\sqrt{z}(1-z)\{(1-\sigma)(1-z)^{-\sigma}+2\kappa \}\: \tilde{\alpha}^2 \:
  \frac{h_{ij}}{l}
\,, 
\end{align}
where $h_{ij}$ and $K_{ij}:=-\nabla_{(i} n_{j)}\,(i,\,j=t,\,\varphi)$ are 
the induced metric and the extrinsic curvature defined on 
a $z={\rm const.}$ surface with $n^\mu$ being the outward pointing unit normal 
vector to it, and then the limit $z\rightarrow 1$ is taken 
in Eq.~(\ref{holographic_tensor}) as a whole. 
Here $\tilde{\alpha}$ is the leading coefficient of the asymptotic 
scalar field $\phi \simeq \tilde{\alpha}(t,\varphi)(1-z)^{\frac{1}{3}}+\kappa \tilde{\alpha}(t,\varphi)(1-z)^{\frac{2}{3}}$.  
Note that the terms in the second line in Eq.~(\ref{holographic_tensor}) 
come from our deformation of the dual field theory by the double-trace interaction.  
Since the divergent term in the first line in $T_{ij}$ cancels with the divergent term in the second line 
in Eq.~(\ref{holographic_tensor}), $T_{ij}$ is well defined on the dual field theory side~(note that $h$ and 
$h_{20}$ in (\ref{Eq:Q}), (\ref{h20}) do not diverge asymptotically). Furthermore, for an arbitrary function 
$\tilde{\alpha}$, it can be easily shown that $T_{ij}$ is conserved by Codacci's equation~\cite{HE}, i.~e.~, $\nabla_i T^{ij}=0$. 
Thus, in this paper, we adopt $T_{ij}$ as the definition of the effective holographic energy momentum 
tensor. Let us define the spatially averaged quantity $\overline{H}$ as $\overline{H}:=\int^{2\pi}_0H(t,\varphi)d\varphi/(2\pi)$. 
Then, at $O(\epsilon^2)$, we find that 
\begin{align}
& {E}^{(2)}_{\rm hair}:=\overline{T^{(2)}}_{tt}\simeq -\frac{h(1)}{2l}\epsilon^2+\frac{r_+^2+r_-^2}{r_+l^3}\epsilon^2 a_1, \nonumber \\
& {J}^{(2)}_{\rm hair}:=-\overline{T^{(2)}_{t\varphi}}=-\epsilon^2\left(\frac{r_+^2-r_-^2}{l}U'(1)-\frac{2a_1r_-}{l^2}\right).  
\end{align}
Combining with (\ref{h_asymp}) yields the first law of the hairy black hole; 
\begin{align} 
{E}^{(2)}_{\rm hair}=TS^{(2)}_{\rm hair}+\Omega_+{J}^{(2)}_{\rm hair}, \qquad 
T:=\frac{r_+^2-r_-^2}{2\pi r_+ l^2}, 
\end{align}
where $T$ is the temperature and $S^{(2)}_{\rm hair}$ is the deviation of the entropy of the black hole given by 
$S^{(2)}_{\rm hair}=2\pi a_1 \epsilon^2/l$. 

Now, let us derive the fourth order corrections of the energy 
${E}^{(4)}_{\rm hair}$ and compare with the energy of BTZ black hole with the same angular momentum and entropy. For the outer~(inner) horizon radius 
$R_+$ and $R_-$, the energy, the angular momentum, and the entropy of the BTZ black hole are 
\begin{align}
\label{BTZ_energy}
E_{\rm btz}=\frac{R_+^2+R_-^2}{2l^3}, 
{J}_{\rm btz}=\frac{R_+R_-}{l^2}, S_{\rm btz}=\frac{2\pi}{l}R_+. 
\end{align}
Let us expand $R_\pm$ as  
$R_+=r_{+}+a_1\epsilon^2+a_2\epsilon^4+\cdots$ and $R_-=r_{-}+r_{1-}\epsilon^2+r_{2-}\epsilon^4+\cdots$  
so that the entropy of the BTZ is equal to the one of the hairy BH, $S_{\rm btz} = S_{\rm hair}$. 
Equating $E^{(2)}_{\rm btz}=E^{(2)}_{\rm hair}$, and ${J}^{(4)}_{\rm btz} = {J}^{(4)}_{\rm hair}$, 
$r_{1-}$ and $r_{2-}$ are expressed by $a_1$, $a_2$, $E^{(2)}_{\rm hair}$, and ${J}^{(4)}_{\rm hair}$, and 
we obtain
\begin{align}
& E^{(4)}_{\rm btz}=\frac{l^3(E^{(2)}_{\rm hair})^2}{2r_{-}^2}
            -\frac{\epsilon^2 a_1(r_{+}^2+r_{-}^2)}{r_{+}r_{-}^2}E^{(2)}_{\rm hair}
            +\frac{r_{-} {J}^{(4)}_{\rm hair}}{lr_{+}} 
\nonumber \\
& \,\,\,\quad \qquad +\frac{\epsilon^4 a_1^2}{2l^3}\left(3+\frac{r_{+}^2}{r_{-}^2}  \right) 
  +\frac{\epsilon^4 a_2(r_+^2-r_-^2)}{r_+l^3}. 
\end{align}
Note that $\Delta^{(4)}E:=E^{(4)}_{\rm hair}-E^{(4)}_{\rm btz}$ is independent of the second order deviation parameter 
$a_2$ in (\ref{expansion}). The numerical calculation shows that $\Delta^{(4)}E\simeq -5.8\times 10^2 \cdot \epsilon^4 <0$. 
This implies that, if we increase  $E^{(4)}_{\rm hair}$ so that it becomes equal to  $E^{(4)}_{\rm btz}$, 
then we have the relationship $S^{(4)}_{\rm hair} > S^{(4)}_{\rm btz}$, due to the thermodynamic laws.

\section{Summary and discussions} 
In this paper, aiming at constructing a rotating black hole whose metric 
admits only a single Killing vector field, we have investigated 
three-dimensional AdS black holes with a real scalar field. 
First we showed that under our more general boundary condition, 
the scalar field perturbation shows instabilities. 
Then we have constructed rotating hairy black holes perturbatively 
with respect to the scalar perturbation amplitude, $\epsilon$, 
from the onset of the instability, up to $O(\epsilon^4)$. 
We showed that the entropy of our hairy black hole is larger than the one of 
BTZ, with the same energy and the angular momentum. 
Judging merely from this entropically favored nature, 
one might expect our hairy black hole to be a possible final fate of 
superradiant instability in AdS$_3$. 
However, since our solution is constructed perturbatively, there is 
a possibility that once non-linear effects are fully taken into account, 
some new unstable modes might show up.   
Also we should mention that under our general boundary condition, 
the background AdS$_3$ itself turns out to be unstable:   
One can check that our boundary condition $\kappa =-0.414$ violates 
the linear stability criterion established in~\cite{AIWald04} (see Eq.~(169) 
in that reference), which in the present case ($m^2l^2=-8/9$, $k=0$) reads off 
\begin{align} 
\kappa \geq - \frac{\left|{\Gamma(-1/3)}\right|\Gamma(2/3)^2}{\Gamma(1/3)^3 } = - 0.387 \,.
\end{align} 
In particular, the $k=0$ mode appears to be a dominant unstable mode over 
other non-axisymmetric $k \neq 0$ unstable modes \footnote{
The authors thank Oscar Dias, Jorge Santos, and Benson Way for pointing out 
this instability. 
}. 
As can be seen in Eq.~(\ref{Eq:sec_phi}), the equations of motion for 
higher order perturbations include a source term that consists only of 
higher (odd-number $k\neq 0$) modes. This implies that 
if there is no zero mode in the initial data for our perturbations, 
the zero mode would not be excited by higher order effects. 
Having such an unstable property, our hairy black hole may have some 
interesting applications in the context of holographic superconductor~\cite{HHH}, 
in which the most dangerous $k=0$ mode of our AdS background would be 
irrelevant (or stabilized) due to, e.g., lattice structure. 
It would be interesting to pursue this possibility.

The most striking feature of our solution is that lumpy geometry oscillates periodically and 
co-rotate with the black hole 
accompanied by the non-linear scalar perturbations. Although it is 
not strictly thermodynamically equilibrium, the system never dissipates~\cite{IIM13}, i.e., the entropy is always constant. 
This is in contrast to the solutions of 
Refs.~\cite{Figueras:2012rb,Fischetti:2012vt}, which actually show dissipation. 
In this sense, our solution may be 
viewed as an extension of the time-periodic solutions 
of~\cite{MaliborskiRostworowski2013} to the case with a black hole added. 
It would also be interesting to generalize the present analysis to 
higher dimensions.  

\acknowledgements
We would like to thank Oscar Dias, Gary Horowitz, Robert Mann, Jorge Santos, and Benson Way 
for comments on the manuscript. 
This work was supported in part by
JSPS KAKENHI Grant Number 25800143 (NI), 15K05092 (AI), 23740200 (KM).



\appendix
\section{Solutions for $P$, $R$, and $\phi_{23}$}
Here we give the expressions of $P$ and $R$ in terms of 
$T$, $T'$, $\Pi_1$, and $\Pi_1'$, 
and the solution $\phi_{23}$ of Eq.~(\ref{Eq:sec_phi}). 

\begin{align}
\label{Eq:P2}
& P=\frac{r_+(r_+^2-r_-^2)(r_+^2+3r_-^2+4k^2l^2)z}{r_-l(r_+^2+r_-^2+2k^2l^2)}T' \nonumber \\
&-\frac{r_+r_-(r_+^2-r_-^2)\{(5r_-^2+4k^2l^2)z-r_+^2(2-z) \}}{l(r_+^2+r_-^2+2k^2l^2)(r_+^2-r_-^2z)}T \nonumber \\
&-\frac{(r_+^2-r_-^2)^2(r_+^2m^2-k^2(1-z))}{2(r_+^2+r_-^2+2k^2l^2)(1-z)(r_+^2-r_-^2z)}\Pi_1^2 \nonumber \\
&-\frac{2(r_+^2-r_-^2)^2(r_-^2+2k^2l^2)z}{l^2(r_+^2+r_-^2+2k^2l^2)(r_+^2-r_-^2z)}\Pi_1\Pi_1' \nonumber \\
&+\frac{2r_+^2(r_+^2-r_-^2)^2(1-z)z}{l^2(r_+^2+r_-^2+2k^2l^2)(r_+^2-r_-^2z)}\Pi_1'^2, 
\end{align}
\begin{align}
\label{Eq:R3}
& R= \nonumber \\
&-\frac{l(r_+^2-r_-^2z)z\{(r_+^4+r_-^4)z-2r_+^2(r_-^2(2-z)+2k^2l^2(1-z))   \}}{2r_-r_+(r_+^2-r_-^2)(r_+^2+r_-^2+2k^2l^2)}T'
\nonumber \\
&+\frac{r_-l\{4r_+^4-8r_+^2r_-^2z+((r_+^2+r_-^2)^2+4(r_+^2-r_-^2)k^2l^2)z^2 \}}{2r_+(r_+^2-r_-^2)(r_+^2+r_-^2+2k^2l^2)}T
\nonumber \\
&+\frac{l^2(r_+^2m^2-k^2(1-z))(r_-^2z-r_+^2(2-z))}{4r_+^2(r_+^2+r_-^2+2k^2l^2)(1-z)}\Pi_1^2
\nonumber \\
&+\frac{(r_+^2-r_-^2-2k^2l^2(1-z))z}{r_+^2+r_-^2+2k^2l^2}\Pi_1\Pi_1' \nonumber \\
&+\frac{(1-z)z(r_+^2(2-z)-r_-^2z)}{r_+^2+r_-^2+2k^2l^2}\Pi_1'^2, 
\end{align}
\begin{align}
&\phi_{23}=\frac{1}{\sigma(\hat{\beta}_3-\kappa\hat{\alpha}_3)}\times \nonumber \\
&\left(\int^z_0 {\cal S}_{\phi3}(z')\lambda_2(z')dz'\lambda_1(z)
+\int^1_z {\cal S}_{\phi3}(z')\lambda_1(z') dz'\lambda_2(z) \right), \nonumber \\
& \hat{\alpha}_3=\frac{\Gamma(\sigma)}
{\Gamma\left(\frac{1+\sigma}{2}+\frac{3il}{2r_+} \right)\Gamma\left(\frac{1+\sigma}{2}-\frac{3il}{2r_+} \right)}, 
\nonumber \\
& \hat{\beta}_3=\frac{\Gamma(-\sigma)}
{\Gamma\left(\frac{1-\sigma}{2}-\frac{3il}{2r_+} \right)\Gamma\left(\frac{1-\sigma}{2}+\frac{3il}{2r_+} \right)}, 
\end{align}
where $\lambda_1(z)$ and $\lambda_2(z)$ are homogeneous solutions of ${\cal L}_{3k}\lambda_i=0, \,i=1,2$; 
\begin{align}
& \lambda_1=(1-z)^{\frac{1-\sigma}{2}}F(\alpha_3, \beta_3, \alpha_3+\beta_3;1-z) \nonumber \\
&\qquad +\kappa(1-z)^{\frac{1+\sigma}{2}}F(1-\alpha_3, 1-\beta_3, 2-\alpha_3-\beta_3;1-z), \nonumber \\
& \lambda_2=(1-z)^{\frac{1-\sigma}{2}}F(\alpha_3, \beta_3, 2;z), \nonumber \\
& \alpha_3:=\frac{3ikl(l\Omega_+-1)}{2(r_+-r_-)}+\frac{1-\sigma}{2}, \quad 
\beta_3:=\frac{3ikl(l\Omega_+-1)}{2(r_+-r_-)}-\frac{1-\sigma}{2}. 
\end{align}
Here, $\lambda_1$ is the solution satisfying our more general boundary condition (\ref{kappa}), while $\lambda_2$ is a regular 
solution at the horizon $z=0$.




\begin{thebibliography}{99}
\bibitem{Maldacena:1997re}
  J.~M.~Maldacena,
  Adv.\ Theor.\ Math.\ Phys.\  {\bf 2}, 231 (1998)
  [Int.\ J.\ Theor.\ Phys.\  {\bf 38}, 1113 (1999)]

\bibitem{BizonRost2011}
P.~Bizon and A.~Rostworowski, 
Phys.~Rev.~Lett.{\bf 107} 031102 (2011). 

\bibitem{CDLY2004}
V.~Cardoso, O.~J.~C.~Dias, J.~P.~S.~Lemos, and S.~Yoshida, 
Phys.~Rev.~{\bf D70} 044039 (2004).  

\bibitem{BritoCardosoPani2015}
R.~Brito, V.~Cardoso, and P.~Pani, 
arXiv:1501.06570[gr-qc].   

\bibitem{KLR2006}
H.~K.~Kunduri, J.~Lucietti, and H.~S.~Reall,  
Phys.~Rev.~{\bf D74} 084021 (2006).  

\bibitem{Reall2003}
H.~S.~Reall,  
Phys.~Rev.~{\bf D68} 024024 (2003).  

\bibitem{DiasHorowitzSantos2011}
O.~J.~C.~Dias, G.~T.~Horowitz, and J.~E.~Santos, 
JHEP {\bf 07} 115 (2011). 

\bibitem{SPMM}
S.~Stotyn, M.~Park, P.~L.~McGrath, and R.~B. Mann, 
 Phys.~Rev.~{\bf D 85} 044036 (2012).  

\bibitem{HawkingReall}
S.~W.~Hawking and H.~S.~Reall, 
Phys.~Rev.~{\bf D61} 024014 (1999).  


\bibitem{HerdeiroRadu2014}
C.~A.~R.~Herdeiro and E.~Radu, 
Phys.~Rev.~Lett.~{\bf  112} 221101 (2014).   

\bibitem{Mann1}
S.~Stotyn and R.~B. Mann, 
J.~Phys.~{\bf A 45} 374025 (2012).  

\bibitem{Mann2}
S.~Stotyn, M.~Chanona, and R.~B. Mann, 
Phys.~Rev.~{\bf D89} 044018 (2014).  

\bibitem{BTZ}
M.~Banados, C.~Teitelboim, and J.~Zanelli, 
Phys.~Rev.~Lett. {\bf 69}  1849 (1992).  

\bibitem{Witten2001}
E.~Witten, 
hep-th/0112258.     

\bibitem{IchiSato94}
I.~Ichinose and Y.~Satoh, 
Nucl.~Phys.~{\bf B447}, 340~(1995) 

\bibitem{Handbook}
Handbook of Mathematical Functions, (Dover, New York, 1965) edited by
M. Abramowitz, and I.A. Stegun. 

\bibitem{BF}
P.~Breitenlohner and D.~ Z.~Freedman, 
Annals.~Phys. {\bf 144}  249 (1982). 

\bibitem{DHIS2013}
O.~J.~C.~Dias, G.~T.~Horowitz, N.~Iqbal, and J.~E.~Santos, 
JHEP {\bf 1404} 096 (2014).  

\bibitem{Birmingham}
D.~Birmingham, 
Phys.~Rev.~{\bf D64}, 064024 (2001). 

\bibitem{StotynMann}
S.~Stotyn and R.~B.~Mann, arXiv:1203.0214[gr-qc]. 

\bibitem{BalaKraus1999}
V.~Balasubramanian and P.~Kraus, 
Commun.~Math.~Phys. {\bf 208} 413~(1999).  

\bibitem{HE}  
S.~W.~Hawking and G.~F.~R.~Ellis, ``The large scale structure of 
space-time'', Cambridge university Press, 1973. 

\bibitem{AIWald04}
A.~Ishibashi and R.~M.~Wald, 
Class.\ Quant.\ Grav.\  {\bf 21}, 2981-3014 (2004)

\bibitem{HHH}
S.~A.~Hartnoll, C.~P.~Herzog, and G.~T.~Horowitz,  
Phys.\ Rev.\ Lett.\  {\bf 101}, 031601 (2008)

\bibitem{IIM13}
N.~Iizuka, A.~Ishibashi, and K.~Maeda, 
Phys.\ Rev.\ Lett.\  {\bf 113}, 011601 (2013)

  
\bibitem{Figueras:2012rb} 
  P.~Figueras and T.~Wiseman,
  Phys.\ Rev.\ Lett.\  {\bf 110}, 171602 (2013)

\bibitem{Fischetti:2012vt} 
  S.~Fischetti, D.~Marolf and J.~E.~Santos,
  Class.\ Quant.\ Grav.\  {\bf 30}, 075001 (2013)

\bibitem{MaliborskiRostworowski2013}
M.~Maliborski and A.~Rostworowski, 
Phys.~Rev.~Lett. {\bf 111} 051102 (2013).      
%


\end{thebibliography}
\end{document}